\documentclass[12pt]{article}

\usepackage{epsfig}

\usepackage{amssymb}
\usepackage{amsfonts}

\usepackage{amsmath}

\def\Journal#1#2#3#4{{#1} {\bf #2}, #3 (#4)}

\def\NPB{{\em Nucl. Phys.} B}
\def\PLB{{\em Phys. Lett.}  B}

\def\PRD{{\em Phys. Rev.} D}

\newskip\humongous \humongous=0pt plus 1000pt minus 1000pt

\newif\ifdtup

\catcode`\@=11

\@addtoreset{equation}{section}
\def\theequation{\thesection\arabic{equation}}

\def\@normalsize{\@setsize\normalsize{15pt}\xiipt\@xiipt
\abovedisplayskip 14pt plus3pt minus3pt%
\belowdisplayskip \abovedisplayskip
\abovedisplayshortskip \z@ plus3pt%
\belowdisplayshortskip 7pt plus3.5pt minus0pt}

\def\small{\@setsize\small{13.6pt}\xipt\@xipt
\abovedisplayskip 13pt plus3pt minus3pt%
\belowdisplayskip \abovedisplayskip
\abovedisplayshortskip \z@ plus3pt%
\belowdisplayshortskip 7pt plus3.5pt minus0pt
\def\@listi{\parsep 4.5pt plus 2pt minus 1pt
\itemsep \parsep
\topsep 9pt plus 3pt minus 3pt}}

\relax

\catcode`@=12

\catcode`\@=11

\def\section{\@startsection{section}{1}{\z@}{3.5ex plus 1ex minus
.2ex}{2.3ex plus .2ex}{\large\bf}}

\def\thesection{\arabic{section}.}

\def\appendix{\setcounter{section}{0}
\def\thesection{Appendix \Alph{section}:}
\def\theequation{\Alph{section}.\arabic{equation}}}

\newcommand{\be}{\begin{equation}}
\newcommand{\ee}{\end{equation}}
\newcommand{\bqa}{\begin{eqnarray}}
\newcommand{\eea}{\end{eqnarray}}
\newcommand{\beas}{\begin{eqnarray*}}
\newcommand{\eeas}{\end{eqnarray*}}

\newcommand{\bquo}{\begin{quote}}
\newcommand{\enqu}{\end{quote}}

\begin{document}

\begin{titlepage}
\begin{flushright}
  IFUP-TH/2005-32  
  \end{flushright}
\bigskip
\bigskip

\begin{center}

{\large  {\bf
DUALITY AND CONFINEMENT IN $SO(N)$ GAUGE THEORIES\footnote{Contribution to Festschrift ``Sense of Beauty in Physics'',
 in honor of Adriano Di Giacomo on his 70th birthday.}
 } }
\end{center}



\bigskip
\begin{center}
{\large
  Luca FERRETTI$^{(1)}$  and   Kenichi KONISHI $^{(2,3)}$  }
 \vskip 0.10cm

\end{center}

\begin{center}
{\it      \footnotesize
SISSA, via Beirut 4,   I-34014 Trieste, Italy $^{(1)}$, \\
Dipartimento di Fisica ``E. Fermi" -- Universit\`a di Pisa $^{(2)}$, \\
Istituto Nazionale di Fisica Nucleare -- Sezione di Pisa $^{(3)}$, \\
Largo Pontecorvo, 3, Ed. C, 56127 Pisa,  Italy  } 

 \smallskip

 { \small  \texttt   l.ferretti(at)sns.it,  konishi(at)df.unipi.it }

\end {center}

\noindent
\begin{center} {\bf Abstract} \end{center}

{The presence of magnetic monopole  like excitations of nonabelian varieties is one of the subtlest consequences of spontaneously broken gauge symmetries.  Important hints about  their quantum mechanical properties, which remained long mysterious,  are coming  from a detailed knowledge   of the dynamics of supersymmetric gauge theories which has become available recently.  These developments might shed light on the problem of confinement and dynamical symmetry breaking in QCD.  We discuss here some beautiful features of vortex-monopole  systems, in  which dual nonabelian transformations among monopoles are generated by the nonabelian vortex moduli.  
  }

\vfill

\begin{flushright}
\today
\end{flushright}

\end{titlepage}

\section{Dual groups from monopole-vortex systems}  

Consider a system with a hierarchical  gauge symmetry breaking  (with  $v_{1}= | \langle \phi_{1} \rangle  |   \gg
v_{2} =| \langle \phi_{2} \rangle  | $) 
\begin{equation}     G   \,\,\,{\stackrel { v_{1}  \ne 0} {\longrightarrow}}     \,\,\, H  \,\,\,{\stackrel {v_{2}   \ne 0} {\longrightarrow}}    \,\, \emptyset,
\end{equation} 
where $ \phi_{1} $ and $\phi_{2}$  are some (elementary or composite)  scalar fields in the theory.
The high-energy theory (at scales much higher than  $v_{2}$)  possesses  regular monopoles, associated with  nontrivial elements of  $ \pi_{2}(G/H)$:  when the unbroken group $H$ is nonabelian, the monopoles will carry some nonabelian gauge charges, generalizing nontrivially the 't Hooft-Polyakov monopole solutions~\cite{TP}, as has been shown by Goddard-Nuyts-Olive, Bais, and others~\cite{NAmonop}.  In fact, 
they are believed to form a multiplet of  a group ${\tilde H}$ which is dual \footnote{The dual group is defined as the group generated by the dual root vectors, $
\alpha^{*} = \alpha / (\alpha \cdot \alpha)$, where $\alpha$ are the nonzero roots of $H$.}   to $H$.  Their fully quantum mechanical properties, however,   have long remained  mysterious.
  
   The low-energy theory with  the gauge symmetry breaking $ H  \,\,\,{\stackrel {v_{2}   \ne 0} {\longrightarrow}}    \,\, \emptyset$
(defined at mass scales much lower  than  $v_{1}$),   has vortices associated with the fundamental group
 $\pi_{1}(H).$    Existence of  the vortices  with exact nonabelian continuous moduli (nonabelian vortices)  in models with an  unbroken 
 color-flavor  type symmetry,  have recently attracted considerable attention~\cite{HT,ABEKY,NAVortex,Eto}. 
 
As $\pi_{2}(G)=\emptyset$ (valid for all Lie groups)   no regular monopole associated with the symmetry breaking $G \to H$  is  truly  stable,  once the smaller vacuum expectation value (VEV)  ($ v_{2} $)  is taken into account ({\it i.e.}, in the full theory).  It means that  they are  confined by the vortex of the low-energy theory.    
On the other hand,   if  $\pi_{1}(G)= \emptyset$ none of  the vortices visible in the low-energy approximation are actually stable in the full theory, either.  In fact, these vortices can be  cut  by  the heavy monopole pair  production, even though this process will be suppressed for $v_{1} \gg v_{2}$. The two phenomena are, actually, the two sides of a medal. 

If $\pi_{1}(G) \ne \emptyset$, however,  there are  some vortices which are stable in the full theory.  These are sourced by, so would confine, the singular Dirac monopoles once they are introduced in the theory.   

One sees that these are questions closely parallel to the general idea of confinement in QCD~\cite{TH};  these models can be regarded as concrete  (dual)  models in which many questions about confinement and dynamical symmetry breaking can be investigated.

For instance, a crucial question about confinement is whether or not it is accompanied by dynamical abelianization. 
Although some of the popular models for confinement in QCD  do involve an abelian effective description of the system,  there are general indications that in Nature and in QCD confinement is not accompanied by dynamical abelianization. 
The fact that in a wide class of supersymmetric theories confinement is described as a dual superconductor but {\it of nonabelian 
type}, makes a more careful investigation of these theories urgent.  A nonabelian superconductor is characterized by vortices which carry continuous, nonabelian moduli of flux. 

Recently  such nonabelian BPS vortices,  {\it i.e.},  BPS vortices  possessing   continuous, exact zeromodes,   have been explicitly constructed in the context of softly broken ${\cal N}=2$ gauge theories, and in related models.
The presence of an exact symmetry, typically a color-flavor diagonal symmetry $H_{C+F}$,  appears to be  fundamental.   
The cases with  $G=SU(N+1)$,  $H=  SU(N) \times U(1) /  {\mathbf Z}_N$  have been  studied in some detail in the papers \cite{HT,ABEKY,NAVortex,Eto}. \footnote{Actually,
in a parallel development \cite{HT,NAVortex,Eto}  only the system $H  \,\,\,{\stackrel {v_{2}   \ne 0} {\longrightarrow}}    \emptyset $  is considered,  with $H=  U(N)$, and with a Fayet-Iliopoulos term in the $U(1)\subset U(N)$ part. The main interest there is the dynamics of vortex orientation modes, which turns out to be various kind of two dimensional sigma models, and their relation to the dynamics of the parent 4D theory.   Monopoles discussed in these papers are abelian monopoles of certain $U(1) $ factors $\subset H$   (kinks of the sigma model).}
   A particularly intriguing idea~\cite{ABEKY} emerging from these studies is that part of these  zeromodes,  due to an exact global symmetry of the system,  are related  to the dual nonabelian transformations among the monopoles of  $\pi_{2}(G/H)$.   
   
   It is the main aim of this note  to pursue  this idea further.  We shall study  in particular some of the questions which arose originally with nonabelian monopoles of  $SO(N)$ theories, but which are of  more general nature,  related to  how the dual (magnetic) group transformations can be understood from the properties of joint monopole-vortex systems in the original (electric) theory.

\section {Vortex moduli and monopoles in  $SU(N+1)\to  U(N) \to \emptyset$}  

Our model~\cite{ABEKY}  was based on softly broken ${\cal N}=2,$  $SU(N+1)$ theory, with symmetry breaking pattern
\begin{equation}  SU(N+1)    \,\,\,{\stackrel {\langle \phi_{1} \rangle    \ne 0} {\longrightarrow}}     \,\,\, U(N) \,\,\,{\stackrel {\langle \phi_{2} \rangle    \ne 0} {\longrightarrow}}    \,\, \emptyset.
\end{equation} 
The first breaking is due to the VEV of a scalar field in the adjoint representation, while the second, smaller VEV is the condensate of the $N_{f}$  squark fields, in the fundamental representation of $SU(N+1)$.   The number of the flavors  $N_{F}$ is such that the $SU(N)$ gauge interactions  do not grow strong  between the scales $\langle \phi_{1} \rangle $ and $\langle \phi_{2} \rangle$:       $2\, N \le N_{f} \le 2\, N+1$.  

\begin{figure}
\begin{center}
\includegraphics[width=2in]{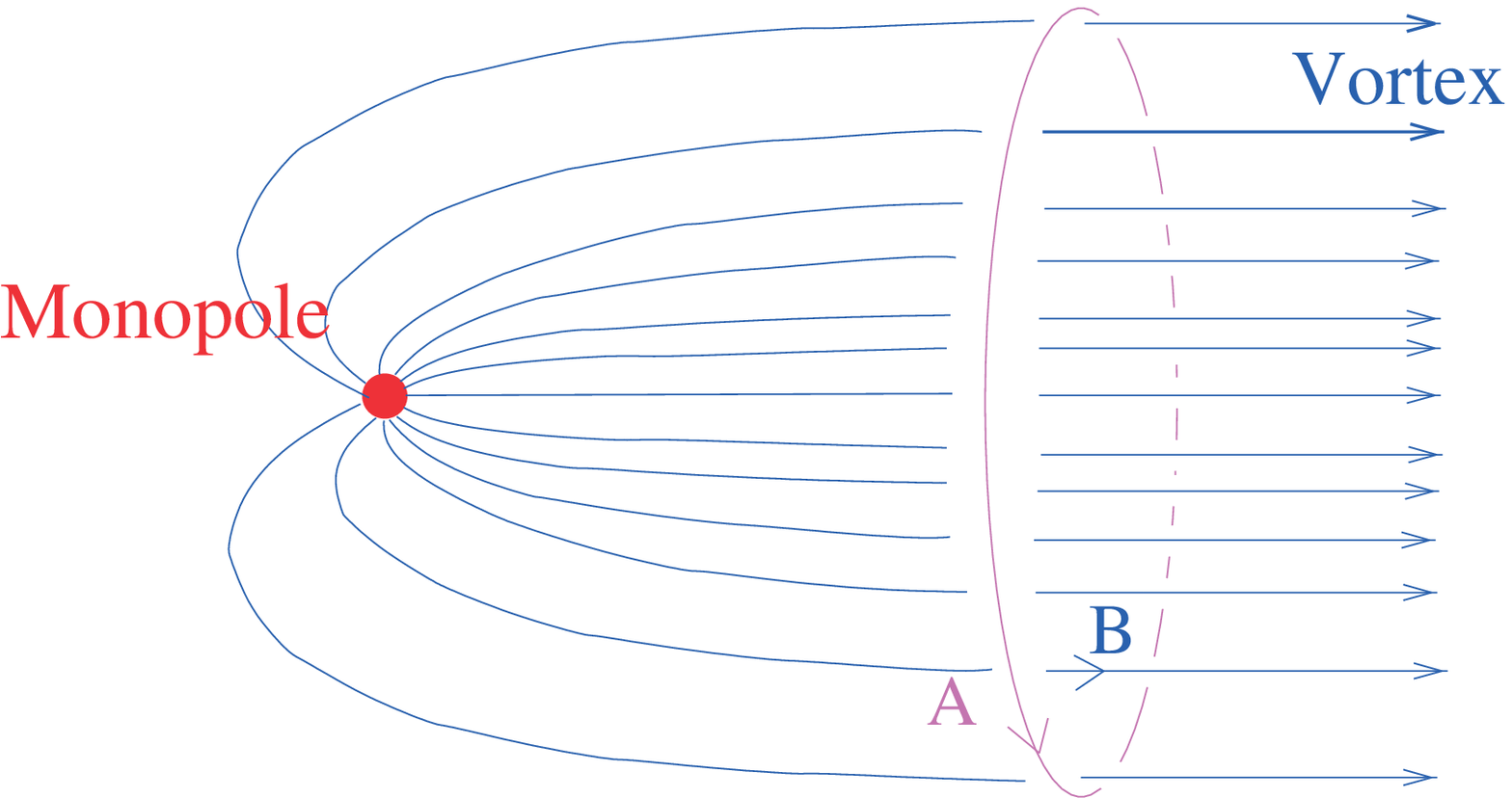}
\caption{ }
\label{monovortex}
\end{center}
\end{figure}

We shall  repeat neither the construction of semiclassical monopoles of high-energy theory, nor the derivation and numerical solution of 
nonabelian Bogomolnyi equations for the low-energy vortices~\cite{ABEKM,ABEKY}.  We limit ourselves to a few   remarks here.   
The crucial feature is that although the gauge symmetry is completely broken in low-energy limit by the squark VEVs, 
a color-flavor diagonal symmetry remains unbroken.    A vortex configuration breaks this symmetry ($SU(N) \to SU(N-1) \times U(1)$), 
generating an exact, continuous vortex moduli (zeromodes).   For the minimum vortex the moduli  is  $CP^{N-1}$, besides the translational modes \footnote{This result is valid for $N_{f}= N_{c}=N$.  For $N_{f}$  larger than $N_{c}$, as in our case, the vortex moduli is actually larger,  having the modes related to the so-called semilocal strings.  We believe however that only this part of the moduli is 
 closely connected to the dual transformations of the monopoles.}.  Note that 
it coincides  precisely with the space  of states   of  a quantum mechanical $N$-state system.  The absence of the monopoles in the full theory is 
accounted for by the fact that the monopole flux can be exactly carried away by a minimum vortex (Fig.~\ref{monovortex}).

 The main idea is that the dual transformations among the monopoles can be seen, in the coupled monopole-vortex system, as the nonlocal $SU(N)_{C+F}$ transformation of the combined infinitely extended monopole vortex configuration.  As it turns out,  in the softly broken ${\cal N}=2$ model studied in Auzzi et.al.\cite{ABEKY},  the approximate semiclassical monopole configuration (nontrivial configuration of the adjoint scalar and gauge fields, with the  squark fields  fixed to their VEVs),  transforms nontrivially only under the $SU(N)_{C}$.  Nevertheless, the whole monopole-vortex configuration (Fig.\ref{monovortex})  is transformed under the flavor group as well, since the low-energy vortex solution involves nontrivial squark configurations.  This means that one is indeed dealing with transformations among physically  distinct objects.  The fact that the vortex moduli is given by $CP^{N-1}$ implies  that the monopoles of minimum charge transform  as a fundamental multiplet  (${\underline N}$)   of the   
  dual $SU(N)$ group.

 The joint,  monopoles-vortex configuration  (Fig.~\ref{monovortex})  is  neither  topologically stable nor BPS saturated.  The monopoles and vortices are  only approximately so, under the condition,   $v_{1} \gg v_{2}. $   The existence of certain parameters upon which the theory depends in a holomorphic way, which allows us to study  the properties of the  monopoles and vortices separately in appropriate effective theories,  are one of the useful  features of supersymmetric systems.

   In the simplest case with the  symmetry breaking, $$ SU(3)    \,\,\,{\stackrel {\langle \phi_{1} \rangle    \ne 0} {\longrightarrow}}     \,\,\, U(2) \,\,\,{\stackrel {\langle \phi_{2} \rangle    \ne 0} {\longrightarrow}}    \,\, \emptyset,$$
  the moduli space of the minimum vortices is  $SU(2)/U(1) \sim S^{2} = CP^{1}$, and is parametr\-ized by two  Euler angles. In other words the  energy of the whole  configuration (Fig. 1)  is  invariant under color-flavor diagonal transformations of the form, $U_{C+F}= e^{i \tau_{2} \beta}\, e^{i \tau_{3} \alpha }$.  The degenerate  $(1,0)$ and $(0,1)$  
  monopoles are transformed to each other by these transformations, so do the $(1,0)$ and $(0,1)$ vortices. 
  
  An apparent problem arises when one considers the monopoles and  vortices of higher winding numbers. It is not difficult to see that the vortices $(2,0)$ and $(0,2)$ are indeed connected by some   $U_{C+F}$ transformations but that no  $U_{C+F}$ transformations exist which connect, for instance,  $(2,0)$ to $(1,1)$ vortex \footnote{We thank W. Vinci for this observation~\cite{Eto,ASY}.}.   What is going wrong? Is the $SU_{C+F}(2)$ group unrelated to the dual $SU(2)$, after all?    What this shows actually  is the fact that the action of the {\it dual} $SU(2)$  group  is not so simply related to the  electric $SU(2)$ group.  The vortices of winding number two  would confine the monopoles of charge two, either in a triplet or a singlet of (dual) 
  $SU(2)$.  The singlet does not transform.  As for the triplet monopoles, the space of the quantum states of a three-state system is $CP^{2}$.  
  It is interesting  that the moduli of the co-axis vortex of winding number two in an $SU(2)$ theory, has recently been studied~\cite{ASY}: it is given by  $CP^{2}$.  This  appears to provide for a further consistency of our picture.

\section {Monopole-vortex system in $SO(N)$  theories}

\subsection{Monopoles}

For concreteness,  we consider here the case of $SO(N)$ theory with odd $N$. 
Let us consider first the breaking pattern,
\begin{equation}  SO(2N+3)    \,\,\,{\stackrel {\langle \phi_{1} \rangle    \ne 0} {\longrightarrow}}     \,\,\, SO(2N+1)\times U(1) \,\,\,{\stackrel {\langle \phi_{2} \rangle    \ne 0} {\longrightarrow}}    \,\, \emptyset. \label{SONbreak}
\end{equation} 
which can be realized, for instance by an adjoint scalar field acquiring a vacuum expectation value (VEV) of the form, 
\begin{equation}   \langle \phi_{1}  \rangle \,=   v_{1} \, \left(\begin{array}{ccc}0 & i & {\bf 0} \\-i & 0 & \vdots  \\{\bf 0} & \ldots  & {\bf 0} \end{array}\right),
\end{equation} 
and by the (much smaller)  squark VEVs discussed below. 
  To find the monopoles, choose $N$   $SO(4)\sim SU(2)\times SU(2)$ subgroups, living in  $(1,2,2i+1,2i+2)$ spaces, $i=1,2,\ldots, N$.  
  They are all broken to $U(1)\times U(1)$  by the VEV above. 
The  $2N$ nonabelian monopoles  can be constructed by embedding the 't Hooft-Polyakov monopoles in these $2N$  $SU(2)$ 
subgroups broken to $U(1)$'s.   One finds thus a natural candidate for the set of monopoles, to  be transformed as the fundamental multiplet of a dual  $USp(2N)$ group.  In order to construct the transformations among them, we make full use of the unbroken $SO(2N + 1)$ group.  The idea is to make a map between the $SO(2N+1)$ generators (antisymmetric matrices) and the $USp(2N)$ generators which have the form, 
\begin{equation}
          \left( \begin{array}{cc}
                  B & A \\ A^{*} & -B^{t}
          \end{array}
          \right),
\end{equation}
where  $A,B$ are $N\times N$ matrices with the constraints, $A^{t}=A$,
$B^{\dagger}=B$,  in an appropriate basis of monopoles.    The  $i$th   $SO(4)\sim SU(2)\times SU(2)$ subgroup is  generated by
(with a simplified notation  $(1,2,3,4) \equiv (1,2,2i+1,2i+2)$)
\begin{equation}    T_{1}^{\pm} = - \frac{i}{2}  \, (\Sigma_{2 3} \pm \Sigma_{ 4 1}), \quad T_{2}^{\pm} = - \frac{i}{2}  \, (\Sigma_{ 3 1} \pm \Sigma_{ 4 2}), \quad T_{3}^{\pm} = - \frac{i}{2}  \, (\Sigma_{12} \pm \Sigma_{4 3}). 
\end{equation} 
The two monopoles  from this $SO(4)$ group are taken to be  $i$-th and $N+i$-th components of  the fundamental representation of 
$USp(2N)$ group. The pair can be transformed to each other by rotations  in the $(2i+2, 2N+3)$
plane ($\subset SO(2N+1)$), thus
\begin{equation}  \Sigma_{2i+2, 2N+3}   \longrightarrow    A_{i, i}.
\label{map1}\end{equation} 
On the other hand,  the two monopoles associated with subgroups $T^{\pm}$ living in the $(1,2,2i+1,2i+2)$ subspace and those living in the 
$(1,2,2j+1,2j+2)$ subspace, $j\ne i$, are transformed into each other  by rotations  in the
 $(2i+1, 2i+2, 2j+1, 2j+2)$ space: they transform in $SO(2N)$ (in the subspace $i=3,4,\ldots, 2N+2$).  In order to see that they actually transform as a pair of 
$U(N)$ representations,  let us  go to the bases    where a $SO(2N)$ vector
naturally breaks to $N+\bar{N}$ under $U(N)$.  This can be done by going from the original $SO(2N)$ basis,\begin{equation}
          \left( \begin{array}{cc}
                  E & F \\ -F^{t} & D
          \end{array}
          \right),
\end{equation}
where $D$, $E$, $F$ are all purely imaginary $N \times N$ matrices  with
the constraints $E^{t} = -E$, $D^{t} = -D$,  to a new basis by 
\begin{eqnarray}
          \lefteqn{
          \left( \begin{array}{cc}
                  1/\sqrt{2} & -i/\sqrt{2}\\
                  -i/\sqrt{2} & 1/\sqrt{2}
          \end{array}
          \right)
          \left( \begin{array}{cc}
                  E & F \\ -F^{t} & D
          \end{array}
          \right)
          \left( \begin{array}{cc}
                  1/\sqrt{2} & i/\sqrt{2}\\
                  i/\sqrt{2} & 1/\sqrt{2}
          \end{array}
          \right) } \nonumber \\
          & &
          = \frac{1}{2}
          \left( \begin{array}{cc}
                  (E+D) + i(F+F^{t}) & i(E-D)+(F-F^{t})\\
                  -i(E-D)+(F-F^{t}) & (E+D) -i(F+F^{t})
          \end{array}
          \right), 
\end{eqnarray}
that is, 
\begin{equation} 
{\psi}_a^{i}=\frac{1}{\sqrt{2}}(\hat\psi_a^{2i-1}+\hat\psi_a^{2i}), \quad
{\psi}_a^{N+i}=\frac{1}{i\sqrt{2}}(\hat\psi_a^{2i-1}-\hat\psi_a^{2i}). \qquad
(i=1,2,\ldots N).
\label{hatbasis}\end{equation} 
It can then be seen that the $U(N)$ elements (acting on  ${\psi}_a^{i}$, $ {\psi}_a^{N+i}$)  are generated by the set of  $SO(2N)$ 
infinitesimal transformations with  $E=D$, $F=  F^{t}$, so 
\begin{equation}    (\Sigma_{2i, 2i} +\Sigma_{2i-1, 2j-1}), \, i \,  ( \Sigma_{2i, 2j-1} - \Sigma_{2i-1, 2j}) \longrightarrow    B_{i, j}. \label{map2}
\end{equation} 
Nondiagonal elements of $A_{ij}$ can be generated by combining the actions of  (\ref{map1}) and (\ref{map2}).

\subsection{Vortices}   

Let us now consider the low-energy theory, with the symmetry breaking 
\begin{equation}   SO(2N+1)\times U(1) \,\,\,{\stackrel {\langle \phi_{2} \rangle    \ne 0} {\longrightarrow}}    \,\, \emptyset.
\end{equation} 
As $\pi_{1}( SO(2N+1)\times U(1)) = {\mathbf Z}_{2} \times {\mathbf Z}$, there are two types of vortices in this theory:  a vortex carrying a ${\mathbf Z}_{2}$ flux  and the $U(1)$  vortices  with a  ${\mathbf Z}$ flux. 
A vortex carrying the minimum flux of one type only, corresponds to  $\pi_{1}(2N+3)={\mathbf Z}_{2}$;  it confines the singular, Dirac monopole of the fundamental theory, if it is introduced in the theory. 

The regular monopoles of minimum charge considered in the preceding subsection are confined by a double-vortex, with the minimum flux  with respect to both factors in ${\mathbf Z}_{2} \times {\mathbf Z}$.  In analogy with the case of $U(N)$  models discussed before,  we wish to construct a model in which an unbroken global $SO(2N+1)_{C+F}$  group  remains,    broken only by individual  vortex configurations, so that continuous zeromodes (moduli) develop.

 It turns out that it is not as straightforward as in the $SU(N)$ case  to construct such a model, with a hierarchical  gauge symmetry breaking 
 (\ref{SONbreak}), 
  in the context of softly broken ${\cal N}=2$  supersymmetric gauge theories, with the superpotential, 
\begin{equation}  {\cal W} =    \{\sqrt{2} {\tilde Q}_i \Phi Q^i + m_i {\tilde Q}_i Q^i \} 
+ \mu \,{\rm Tr} \, \Phi^2,   \qquad    m_{i} \to m.   
\end{equation} 
  It is however quite easy to  add an ${\cal N}=1$ superpotential
  \begin{equation}      \Delta {\cal W} =  M \, ( Q_{i}\, Q_{i}  +  {\tilde Q}_i \, {\tilde Q}_i ),\qquad  M=m
  \end{equation} 
so that the vacua with desired  properties do appear.  For instance, in the case of the $SO(5)$  theory, the adjoint VEVs can be taken as
\begin{equation}  \phi =   \frac{ 1}{ \sqrt{2}} 
\left(\begin{array}{ccc}  \left(\begin{array}{cc}0 &  i \, m_1 \\ - i\, m_1 & 0\end{array}\right) & {\bf 0}  & 0 \\{\bf 0}  & {\bf 0} & 0 \\0 & 0 & 0\end{array}\right), \qquad  \phi_{1}=  \sqrt{2}   m_1,  
\end{equation} 
which breaks the gauge symmetry as $SO(5) \to SO(3) \times U(1)$, at a high mass scale,  $v_{1}=  m_{1}$, while  the squark VEVs take the form, 
\begin{equation}     Q_1=  \left(\begin{array}{c}d_1 \\-i \, d_1 \\0 \\0 \\0\end{array}\right)
\qquad   {\tilde Q}_1=\left(\begin{array}{c}{\tilde d}_1  \\ i \, {\tilde d}_1 \\0 \\0 \\0\end{array}\right),
 \qquad |{\tilde d}_1|^2= d_1^2 \ll  m_{1}^{2}, \quad {\rm Re} \, d_{1}\,{\tilde d}_1=  - \frac{ \mu \, m_1}{2},
\end{equation} 
and 
\begin{equation}     
 Q_2=\left(\begin{array}{c}0 \\0 \\w \\0 \\0\end{array}\right), \qquad 
  Q_{3}= \left(\begin{array}{c}0 \\0 \\ 0 \\  w \\0\end{array}\right),
 \qquad
   Q_4=\left(\begin{array}{c}0 \\0 \\ 0 \\0 \\ w\end{array}\right),  \label{Qvev1}\end{equation} 
  \begin{equation}   
   {\tilde Q}_2= \left(\begin{array}{c}0 \\0 \\ -w \\0 \\0 \end{array} \right),  \qquad 
    {\tilde Q}_3= \left(\begin{array}{c}0 \\0 \\ 0 \\  -w \\0\end{array} \right),  \qquad  
    {\tilde Q}_4=\left(\begin{array}{c}0 \\0 \\ 0 \\0 \\ -w\end{array} \right).    
\label{Qvev2}  \end{equation}   
The squark VEVs  with  $w, \sqrt{ m_{1} \, \mu} \sim v_{2}  \ll v_{1}$  break the gauge symmetry completely at low energies, leaving  
 an exact $SO(3)_{C+F} $ symmetry. \footnote {Another possibility for constructing a model of this sort, with nonabelian vortex moduli,  is to introduce more than one adjoint scalar fields, $\Phi_{i}$ (L.Ferretti, unpublished).}

\begin{figure}
\begin{center}
\includegraphics[width=4in]{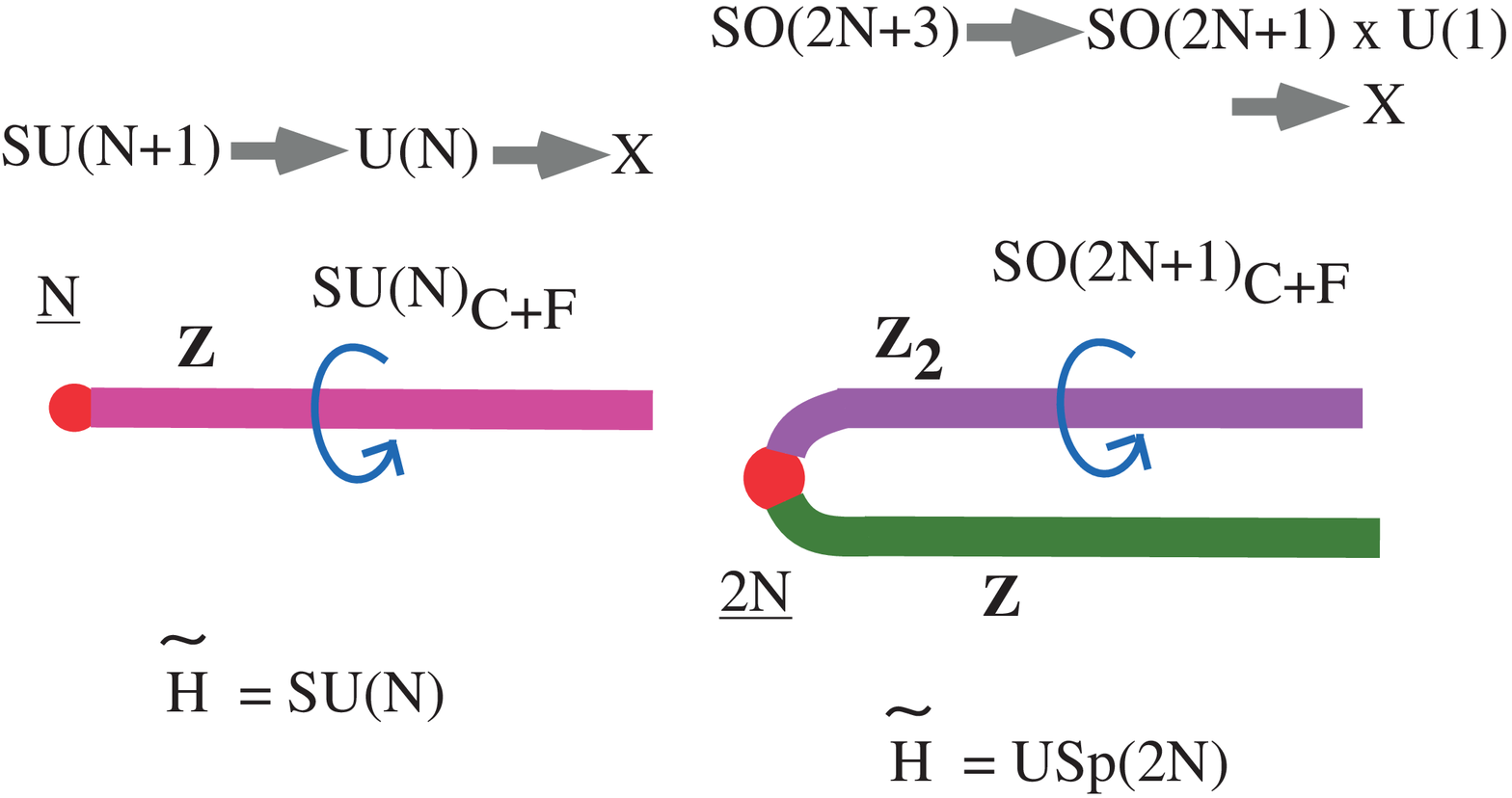}
\caption{ }
\label{sovortex}
\end{center}
\end{figure}

In conclusion, the idea of  nonabelian confinement, with dual group transformations among the monopoles generated by vortex zeromodes (moduli),  can  be naturally generalized to this system,  although in this case the regular monopoles  transform  in the dual, $USp(2N)$ group and the confining string consists of two types of vortices. (Fig.~\ref{sovortex})

\section {$SO(2N+1) \to U(N) \to \emptyset$}

The magnetic monopoles in these cases,  according to the Goddard-Nuyts-Olive construction,  belong to the second rank (symmetric) tensor representation of the dual $SU(N)$  group, and should  transform as such. Again, by constructing the base of degenerate  $\frac {N(N+1)}{2} $  monopoles  following E. Weinberg~\cite{NAmonop},   it is a simple matter to check that not all of these monopoles are connected by the unbroken  $SU(N)_{C}$  transformations.  This is another example of the nontrivial relation between the actions of the electric group $H$  or $H_{C+F}$ under which the monopole-vortex complex transforms,  and the dual group ${\tilde H}$  under which the nonabelian monopoles are believed to  transform.  In the simplest case $N=2$, again the result of Hashimoto et. al. and Auzzi et. al.~\cite{ASY} that the vortex moduli of the winding number two  be  $CP^{2}$,   shows the consistency of our picture, the monopoles being in a triplet in this case.

\section {Conclusion}

Although  some very detailed quantum properties of nonabelian monopoles are known now (for instance, \cite{GKY}),  the precise relation between the dual transformation of the monopoles and the electric, unbroken symmetries,  has yet to be fully elucidated.  We hope that studies along the line of this note will help in clarifying these issues, and eventually in understanding the  confinement itself.

\section*{Acknowledgments}
 The authors thank Naoto Yokoi and Walter 
 Vinci  for discussions.

\end{document}